**Neutron Photcrystallography: Simulation and Experiment**

J. Schefer[1,*], D. Schaniel[2], V.Petříček[3] and Th. Woike[2]

1. Laboratory for Neutron Scattering, ETH Zürich and Paul Scherrer Institut, WHGA-244, CH-5232 Villigen PSI, Switzerland
2. I. Physikalisches Institut, Universität zu Köln, Zülpicherstrasse 77, D-50937 Cologne, Germany
3. Institut of Physics v.v.i., Academy of Sciences of the Czech Republic, Na Slovance 2, 182 21 Praha 8, Czech Republic

Abstract.

The investigation of light-induced structural changes by diffractive methods has improved significantly in the last two decades. We present here the case of neutron photocrystallography for which we have built a special experimental setup at the single crystal neutron diffractometer TriCS at the Swiss Spallation Neutron Source SINQ. We illustrate the progress of the method on the example of the structural determination of photoinduced nitrosyl linkage-isomers in $Na_2[Fe(CN)_5NO]\cdot 2H_2O$. The in-situ determination of the population of the light-induced linkage isomers by optical transmission measurements enhances the reliability of such structural investigations considerably. Additionally we present a new simulation tool within the program package JANA2006 which allows to plan a photocrystallographic experiment thoroughly since the required q-range, the minimally needed population of the photoinduced species, as well as the necessary counting statistics for a successful single crystal diffraction experiment can be evaluated in advance.

[*] Correspondence author (Jurg.Schefer@psi.ch)

# 1. Introduction

Reversible photoinduced chemical reactions or structural changes have received much attention in the last decades due to the possibility to trigger such changes by laser pulses which allows for a targeted investigation of the fundamental underlying processes, e.g. charge transfer processes or isomerisation reactions ([1-5] and references therein). In this manner structures and chemical complexes become accessible which cannot be realized by standard synthetic methods. Static neutron diffraction experiments investigating light-induced structural changes were started on the optically switchable system $Na_2[Fe(CN)_5NO]2H_2O$ (SNP) [6] which is of interest due to its outstanding photorefractive properties [7]. In SNP two metastable linkage-isomers, labelled SI and SII, of the NO-ligand are generated by light irradiation in the blue-green spectral range as illustrated in Fig. 1 [3, 8].

The first structural investigation of the metastable states in SNP was performed by neutron diffraction [6], which found a small increase in the Fe-N bond length. This result was later confirmed by x-ray diffraction [9], but did not satisfactorily explain the long lifetime of the metastable states [10]. In 1997 a breakthrough in the understanding of this kind of metastable states was made by the x-ray diffraction study of Carducci et al. [11] and the calculations based on density functional theory (DFT) by Delley et al. [12], which showed that in SI the NO ligand is rotated by 180° (isonitrosyl) and in SII it is rotated by 90° (side-on configuration). Surprisingly a following neutron diffraction study failed to confirm the isonitrosyl configuration for SI in SNP [13]. Only recently two neutron diffraction studies on the states SII [14] and SI [15] unambiguously confirmed the linkage-isomer character of the metastable states in SNP. For this success a series of experimental improvements were necessary, which will be discussed in this paper.

Today, X-ray diffraction is used for most of the photocrystallographic experiments, but improved neutron diffraction investigations have been proven to be especially suited for the investigation of light-induced linkage isomers due to the following reasons: i) the neutron is a low energy probe, i.e. it does not destroy the light induced species, ii) neutrons probe the nuclear density, i.e. they are not disturbed by additional purely electronic changes of the investigated system, iii) the scattering lengths of nitrogen (9.36 fm) and oxygen (5.803 fm) differ significantly, iv) neutrons are sensitive to light atoms such as hydrogen, v) the differing scattering lengths of isotopes can be exploited to investigate specific problems, and vi) light-induced magnetic ordering can be probed.

A common problem in all photocrystallographic experiments (x-ray or neutron or electron diffraction) is the fact that only a fraction of the molecules (or sample) can be transferred to the excited or metastable configuration, such that one deals with a mixed system of ground state and excited/metastable state in the diffraction experiment. In an improved photocrystallographic setup one needs to determine the amount of excited species in order to exclude this parameter from the refinement.

In the following we describe the state-of the art in neutron photocrystallography, which we achieved after many improvements of the technique such that now it is available to users worldwide at SINQ. The method, the necessary equipment and data treatment as well as the optical calibration experiments to determine the population of excited species is presented on the example of SNP, where neutron diffraction gives direct evidence for the nitrosyl linkage-isomers [14,15].

## 2. Neutron photocrystallography

### 2.1 The method

Experiments have been performed in four steps: (1) The characterization of the optical properties, (2) the measurement of the ground state on the neutron single crystal diffractometer, (3) the in-situ illumination of the crystal and (4) the collection of the neutron data set of the mixed system ground/metastable state. For SNP, we performed such measurements for the two metastable states, SI and SII.

The ground state data set is necessary as a reference (difference Fourier maps). Refinement has been done using JANA2000 [16], now available as JANA2006. The detailed results are described in separate publications (SI [15], SII [14]). We discuss here the special details concerning photocrystallography, e.g. the refinement of mixed states (GS/SI and GS/SII with partial populations $P_i$) with necessary restrictions as well as the results from the support measurements such as absorption α used in order to reduce parameters and to increase the quality of our results. We also discuss in detail the relation between the optical absorption properties and the populations $P_i$.

### 2.2 The neutron diffractometer TriCS

The neutron diffraction experiments on the metastable linkage isomers in SNP were performed on the single-crystal diffractometer TriCS [17], which is located at the Swiss Spallation Neutron Source SINQ [18] at PSI/Villigen. TriCS is a dedicated single crystal neutron diffractometer equipped with 1 single detector and 2 area detectors of 160 by 160 mm$^2$ active area. The detectors are separated by 45$^o$ in 2Θ, and individually tiltable +30/-10$^o$. Two focusing monochromators are available: Ge$_{311}$ delivering a wavelength of 1.180 Å and C$_{002}$ delivering 2.318 Å. The maximum flux reached for 1.180 Å at the sample position of the instrument was 1.04 10$^6$ n/cm$^2$/s using Megapie [19], a liquid Pb/Bi target, and a continuous proton beam of 1.35 mA/0.8 MW on target.

The sample environment of the single crystal instrument TriCS at the Swiss spallation neutron source SINQ has been modified in order to allow in-situ illuminations with up to 200 mW/cm$^2$ in the wavelength range from 400 to 800nm on the instrument directly before the collection of the datasets at temperatures down to 20K, depending on the heat conductivity of the sample. Other wavelength ranges can be realized by exchanging the light source or installing external user equipment.

## 2.3 Sample preparation and illumination

In order to investigate the metastable nitrosyl linkage-isomers by static photocrystallography the samples need to be cooled below ~100K [6]. We therefore installed a closed cycle cooling machine schematically shown in Fig. 2, which delivers 10K under standard operation. As we are using open heat shields to have access with light, but also due to the absorbed light during the illumination process, the lower limit is between 20K and 30K for photocrystallographic applications. During the measurement, the sample has to be kept under vacuum. Many sample such as SNP are not stable under vacuum at room temperature as they contain crystal water. We overcome this limit by undercooling the sample down to approximately 250K, before applying the insulating vacuum. In earlier data collections, both X-ray [11] and neutrons [6,20,21], the population number reached for the metastable state was always a matter of debate. As a consequence, we have chosen in the latest experiments an in-situ illumination process directly on the instrument, which allows us to monitor the population number with the same light as used for the population of the metastable states. This has been reached by installing a fiber optic beyond the crystal as indicated in Figures 2 and 3. This method allows us to measure the transmitted light outside of the closed cycle refrigerator and hence to determine the population of the metastable states with high accuracy.

We cut the SNP crystals normally perpendicular to the orthorhombic *a* axis to plates with a thickness about $d$=0.05 cm a diameter of 0.7 cm. This is a compromise between resolution, population and intensity: Too big crystals lower the resolution significantly, too thick crystals do not allow a homogenous population, and too small crystals lower the intensity of the neutron signal and limit the reliability of the measurement as discussed later in the simulations used for planning the experiments. A flat crystal close to the maximum diameter (limited by a homogenous illumination of the neutron beam) has been chosen according to these constraints. The temperature is kept constant around 50(2) K during data collection. Illumination of the crystal was performed using a Xenon lamp and a set of filters, as indicated in Fig. 3. Filtering the light was reached in several steps: $H_2O$ filter to remove the infrared light, broad band filter $F_1$ to shorten the wavelength band and remove the major part of the unwanted radiation, interference filter $F_2$ for a first selection of the illumination band and the final interference filter $F_3$ to give a narrow wavelength band before polarizing the blue light along the *c* axis of the crystal, the most efficient direction for population as the c-axis is perpendicular to all NO-bonds in the case of SNP. The total exposure was about 5000 J/cm$^2$ using an average power of only 20 mW/cm$^2$. It is known from wavelength-dependent differential scanning calorimetry (DSC) measurements [22] that illumination in the spectral range of 430 to 470 nm with polarization along the *c* axis of the crystal yields a SI population of about 45–50% and a SII population of about 3–5%.

In order to obtain a high population of SII the following procedure has been proven to be most adequate: First the state SI is produced using light in the blue-green spectral range as described above. Then SI is transferred to SII using infrared light [22]. Using the wavelength 1064 nm, about 60% of the SI population is transferred to SII, while the remaining decay back into the ground state, such that SII populations above 20% can be reached by this procedure [25]. The transfer can be monitored by recording the transmitted light intensity of the infrared light. In this manner the population of SII can be determined from the transmission measurement as done in Ref. [14], where 16% of SII were reached.

**2.4 From optical absorption to population**

In SNP, the generation of the metastable linkage isomers is connected with optical absorption changes, which are known from low-temperature absorption spectroscopy [25]. Therefore the change of the transmitted light intensity during illumination can be used to determine the population numbers of the metastable states, *SI* and *SII*. For new systems the absorption curves would have to be determined first.
The used photocrystallographic setup of TriCS@SINQ allows for the in situ measurement of the optical absorption with a Silicon photodiode as described above. In order to have a well-defined signal, the transmitted light was narrowed by an interference filter at 458 nm with a narrow transmission band (full width at half maximum of 2 nm). The detected time-dependent photovoltage $U(t)$ can be described by

$$U(t) = a \cdot I(t) = a \cdot I_o e^{-\alpha(t) \cdot d} \quad (1)$$

where $\alpha$ denotes the absorption coefficient, $d$ the crystal thickness, and $I_0$ the constant incident light intensity. $a$ is a proportionality constant, describing the detection efficiency of the diode. During illumination the absorption changes as a function of exposure $Q=I_0 \cdot t$, where t is the time, due to the change in population of the three involved states GS, SI, and SII. Hence the total absorption coefficient is given as the sum of the single contributions

$$\alpha(t) = \alpha_{GS} \cdot P_{GS}(t) + \alpha_{SI} \cdot P_{SI}(t) + \alpha_{SII} \cdot P_{SII}(t) \quad (2)$$

where $P_{GS,SI,SII}$ denote the populations of the different states GS, SI and SII respectively. It is known from the low-temperature steady-state absorption measurements [25] that $\alpha_{SI}=0$ at 458 nm (wavelength of the illuminating light), so that Eq. 2 reduces to

$$\alpha(t) = \alpha_{GS} \cdot P_{GS}(t) + \alpha_{SII} \cdot P_{SII}(t) \quad (3)$$

Since every molecule has to occupy one state, the sum of $P_{GS}+P_{SI}+P_{SII}$ equals to 1, and we obtain

$$\alpha(t) = \alpha_{GS} - \alpha_{GS} \cdot P_{SI}(t) + (\alpha_{SII} - \alpha_{GS}) \cdot P_{SII}(t) \quad (4)$$

At the beginning of illumination all molecules are in the ground state and hence the photovoltage is

$$U_o = U_{t=0} = a \cdot e^{-\alpha(0) \cdot d} = a \cdot I_o e^{-\alpha_{GS} \cdot d} \quad (5)$$

Then we can write the time-dependent voltage as

$$U(t) = U_o \cdot e^{\alpha_{GS} \cdot P_{SI}(t) \cdot d} \cdot e^{(\alpha_{GS} - \alpha_{SII}) \cdot P_{SII}(t) \cdot d} \quad (6)$$

Equation 6 can be written in the form

$$\ln\left[\frac{U(t)}{U_0}\right] = \alpha_{GS} \cdot P_{SI}(t) \cdot d + (\alpha_{GS} - \alpha_{SII}) \cdot P_{SII}(t) \quad (7)$$

As a function of exposure $Q$, the population of $SI$ and $SII$ is given by

$$P_{SI,SII}(Q) = P^{sat}_{SI,SII} \cdot \left[1 - e^{\frac{-Q}{Q_{SI,SII}}}\right] \quad (8)$$

with the characteristic exposures $Q_{SI}$ and $Q_{SII}$ and $P^{Sat}$ the saturation population of SI and II respectively. Fitting the function

$$\ln\frac{U(t)}{U_o} = A_{SI} \cdot (1 - e^{-\frac{Q}{Q_{SI}}}) + A_{SII} \cdot (1 - e^{-\frac{Q}{Q_{SII}}}) \quad (9)$$

we obtain the parameters $A_{MS}$ and $Q_{MS}$. Since we know the absorption values $\alpha_{GS}=103 cm^{-1}$ and $\alpha_{SII}=158 cm^{-1}$ for our illumination wavelength $\lambda=458$ nm from optical measurements as well as the thickness of the crystal, we can calculate the populations of SI and SII (in our case 40% for SI and 4% for SII [15]), in good agreement with absorption, infrared, and Fe-Mössbauer measurements as described elsewhere [22].

## 2.4 Population determination by neutron diffraction: an alternative for special cases

As an alternative option for the determination of the population of the metastable states, direct measurements of the neutron form factor, F, of a reflection sensitive to the structural changes can be done. In the case of SNP, this is possible, as all structure factors are real (space group Pnnm), a fact which has been already used by others to calibrate time-dependent intensities in the case of the decaying metastable states measured with x-rays [11]. Thereby the structure factor is given by

$$F = n \cdot F_{GS} + (1-n) \cdot F_{MS} = n \cdot \Delta F$$

$$F \approx \sqrt{I} \quad for \quad Pnnm \tag{10}$$

where n is the population of the metastable species.
Fig. 5 shows the results obtained on the reflection (107) for SNP which was calibrated against the optical absorption. The structure factors and the assumed population of earlier neutron diffraction measurements are indicated showing rather good agreement.
The disadvantage of this method is, that the reflections have to be selected by results not measured yet, but just predicted by theoretical calculations such as DFT (for SNP, e.g. [12]) or by assumption from spectroscopic measurements such as Raman (reference [23] for SNP or [24] for $K_2\{RuCl_5)NO\}$. In the case of neutrons, extinction as a result of the heat load from the illumination may influence the results. Furthermore the error is linear in the population n (see Eq.10). As a consequence, we used for the determination of the population in the case of SNP optical absorption measurements rather than this method. However, this method can be an alternative if optical absorption measurements are not possible. The results shown in Fig. 5 show excellent agreement between the two methods in the case of SNP.

## 2.5 Structure refinement of SI and SII in SNP

Also in SNP - as in all other systems known up to date - only a limited population of the metastable states has been achieved. As a consequence, the refinement of the two data sets (SI and SII) had to be done on a mixed system of the ground state and at least one metastable state as explained above. From our absorption measurement [25], we know the population of the crystal after the illumination. In the case of the SI-data-set the values were determined to be 4% of the NO ligands in the *SII* position, 40% in the SI position, and 56% in GS [15], for SII-data set 16% in SII and 84% in GS [14]. We also measured a data set of the pure ground state (GS) for reference. The refinement package used was JANA2000 [16]. In a first step, difference Fourier maps between the ground state structure and the observed data sets SI and SII (photo difference Fourier maps) were calculated. The difference is a result of the photoinduced states. They easily give the location of the atoms involved in the structural change. In a second step we performed refinement with part of the molecule in the ground state and part of the molecule in a structurally different metastable state. Such a refinement is limited by assumptions e.g. that the central Fe atom or the CN-ligands are not involved in the structural change and stay unchanged. It is therefore important to watch temperature factors as wrong assumption may be covered by such corrections.

The structure factor F for any hkl can be written as a function of the structure factors of the ground states GS and the metastable state S (where S is SI or SII) :

$$F = (1 - n_S)F_{GS} + n_S F_S = F_{GS} - n_S \cdot \Delta F$$

*with*

(10)

$$\Delta F = F_{GS} - F_S$$

In the special case of SNP with spacegroup Pnnm, all structure factors are real, and we can write

$$F(t) = \sqrt{I_0(t)}$$

(11)

For X-ray diffraction, the destruction of the metastable states as a function of time (t) has to be taken into account. This can be done by describing the population $n_S(t)$ (S=SI,SII) as

$$n_S(t) = n_{t=0,S} \cdot e^{-\gamma \cdot t}$$

(12)

, as done successfully by Carducci et al. [11], however, complicating the refinement further. Due to the lower energy of the probing neutron, $n_S$ is constant for neutron diffraction.

**2.5 Simulations of photocrystallographic Fourier maps using JANA2000**

In photocrystallography, we are mostly faced to the measurement of mixed states, as a full population can not be reached due to internal limits of the system. Planning the measurement is therefore an essential step. We therefore introduced a simulation mode into JANA2000 (as well as the successor JANA2006) and have tested it successfully on the example of SNP. As our main goals are the photo Fourier maps in the most interesting plane involving the NO molecules, we tested them on the example of SII for three parameters which can vary in the experiment: population of the metastable states SII, coverage of the q-range (max $\sin(\Theta)/\lambda$ we will measure) and statistics (represented by $\sigma$). $\sigma$ can be influenced by introducing artificial noise to the data. Figure 6 shows the different plots illustrating the expected visibility of the side-on configuration of the NO-bond in SII (horizontal NO-bond in the figures) in the presence of the ground state GS (vertical NO-bond in the figure). The results are obvious: The increase of the q-range $\sin(\Theta)/\lambda$ from 0.4 Å$^{-1}$ to 0.7 Å$^{-1}$ is increasing the quality of the data significantly more than increasing the population from 10% to 20%, whereas increasing the statistics has a lower influence. In consequence, we made a measurement on a 20% populated crystal covering 0.7 Å$^{-1}$ in $\sin(\Theta)/\lambda$ on a thin crystal, therefore limiting our counting statistics in order to reach an optimal result. The method successfully applied to SNP can be used for any system, also to standard systems containing only one molecule. In our case, increasing the coverage of $\sin(\Theta)/\lambda$ from 0.4 Å$^{-1}$ to 0.7 Å$^{-1}$ elevates the number of reflections to be collected by a factor of 4.4 (SNP, $\lambda$=1.18 Å), similar to the factor of 4 needed to reduce the statistical error from 6$\sigma$ to 3$\sigma$, but the increase in the q-range is significantly more improving the Fourier maps as is shown by Fig. 6.

## 3. Conclusions

The neutron diffraction experiments on the metastable excited states SI and SII of Sodiumnitroprusside (SNP, Na$_2$[Fe(CN)$_5$NO]·2H$_2$O) using neutron single crystal diffraction show the improvement reached in photocrystallography within the last decade. The in-situ determination of the population of the metastable states by measuring absorption of the illuminating light considerably improved the reliability of the experiments since this important parameter is eliminated from the refinement and replaced by a directly measured value. The simulation of photo difference Fourier maps with JANA2006 [16] allows to determine the conditions for a successful experiment such as the required q-range and minimum population as well as the measurement statistics to reach desired error ($\sigma$) for the diffraction data sets.

[*] Correspondence author (Jurg.Schefer@psi.ch)    page   9

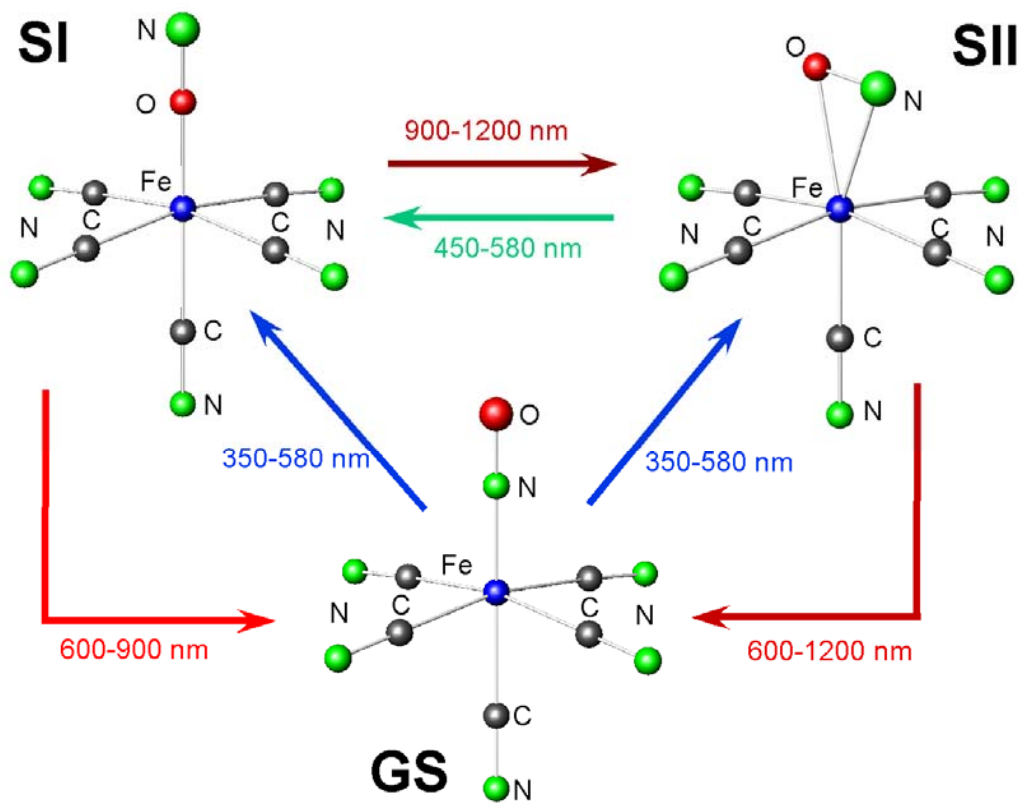

**Fig. 1:** Schematic illustration of the photogeneration of nitrosyl linkage isomers in Na$_2$[Fe(CN)$_5$NO]2H$_2$O (SNP).

* Correspondence author (Jurg.Schefer@psi.ch)  

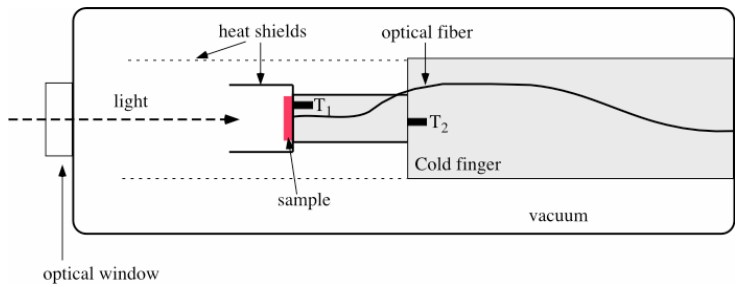 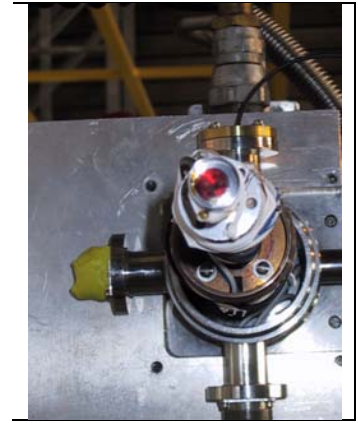

**Fig. 2:** a) Schematic drawing of the cooling device with the sample (red crystal in the on-line version), temperature sensors $T_1$ and $T_2$, one heat-shield at sample temperature and one at liquid Nitrogen temperature (dotted line). The optical window is made of Quartz for our case, but can be exchanged to standard glass or ZnSe if needed. The standard optical fiber behind the sample is transferring the transmitted light out of the device for an in-situ absorption measurement during illumination. b) Detail of the sample holder (top-view to the cooling device). The optical fiber is visible on the upper right hand side and as a black point on the right/lower side of the transparent cylindrical crystal.

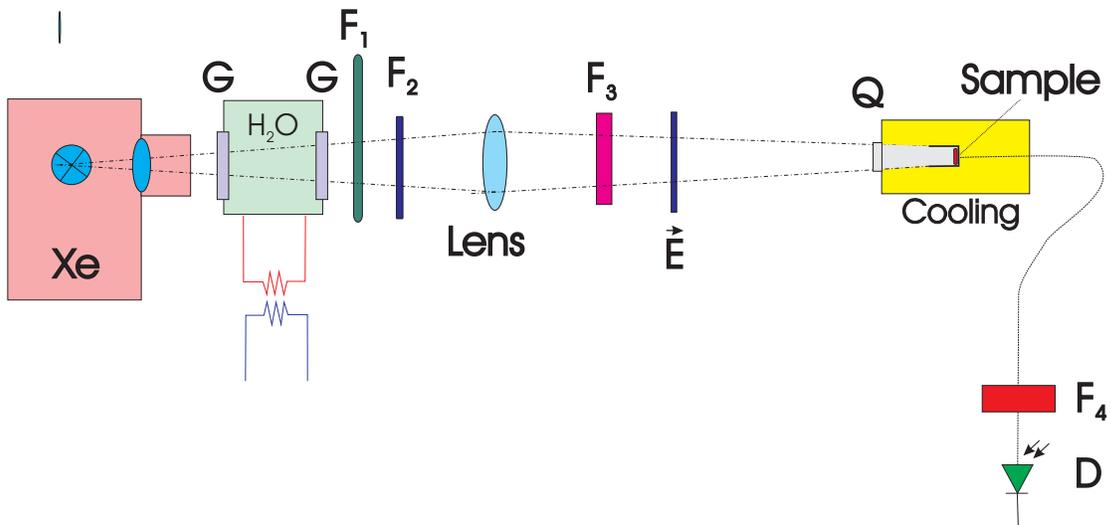

**Fig 3:** Illumination device and filters. The Xenon lamp (Xe) is on the left. Devices Xe, G, $F_1$, $F_2$, Lens, $F_3$ and E are mounted on an optical bench, devices Q, cooling device on the diffractometer, $F_4$ and D in the experimental area.

| | |
|---|---|
| Xe: | ozone-free Xenon lamp, 1000 Watt with integrated lens, Lot-Oriel 6271 |
| $H_2O$: | cooled water filter (demineralized water, 25°C) necessary to filter infrared light |
| G: | standard glass windows (transparent for λ >200nm) |
| $F_1$: | dichroic broad band filter 350-500nm |
| $F_2$: | band pass Filter 400-520 nm |
| Lens: | f=300mm |
| $F_3$: | interference filter 450+/- 20 nm |
| E: | Polarization Filter, E vertical |
| Q: | Quartz-Window (transparent for λ > 300nm) |
| Sample: | See special figure for details |
| $F_4$: | interference filter 458nm +/- 2nm |
| D: | Si-Photodiode |

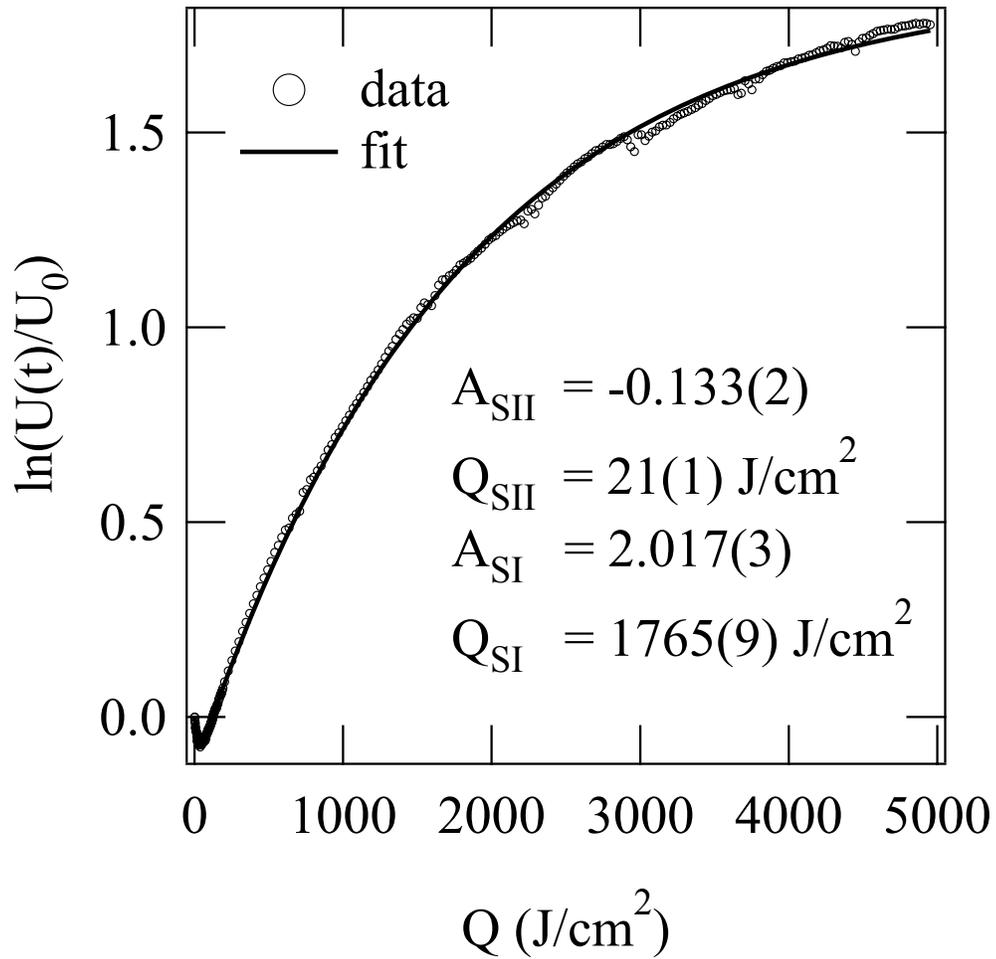

**Fig. 4:** Absorption curve for Sodiumnitroprusside (SNP) measured for λ=458 nm through the optical fiber with the diode D shown in Fig. 3. The first drop in the absorption results from the build-up of SII, before the transfer to SI gets dominant as illustrated in the illumination scheme in Fig. 2.

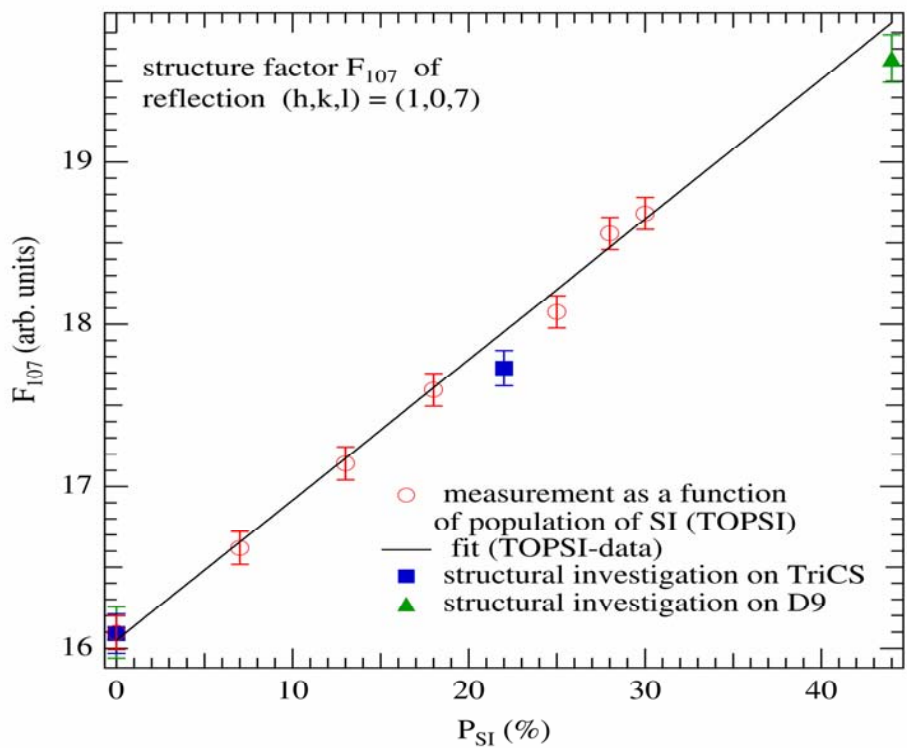

**Fig. 5:** Measurement of the population of SI in Sodiumnitroprusside (SNP) using neutron diffraction on Morpheus (former name: TOPSI). The values are in excellent agreement with the other methods used on TriCS (light absorption, [13]) and D9 (illumination time estimated from previous Mössbauer measurements, [6]).

* Correspondence author (Jurg.Schefer@psi.ch)                                                                                   

**Fig. 6a**

a) $\sin(\Theta)/\lambda$ up to 0.4 Å$^{-1}$

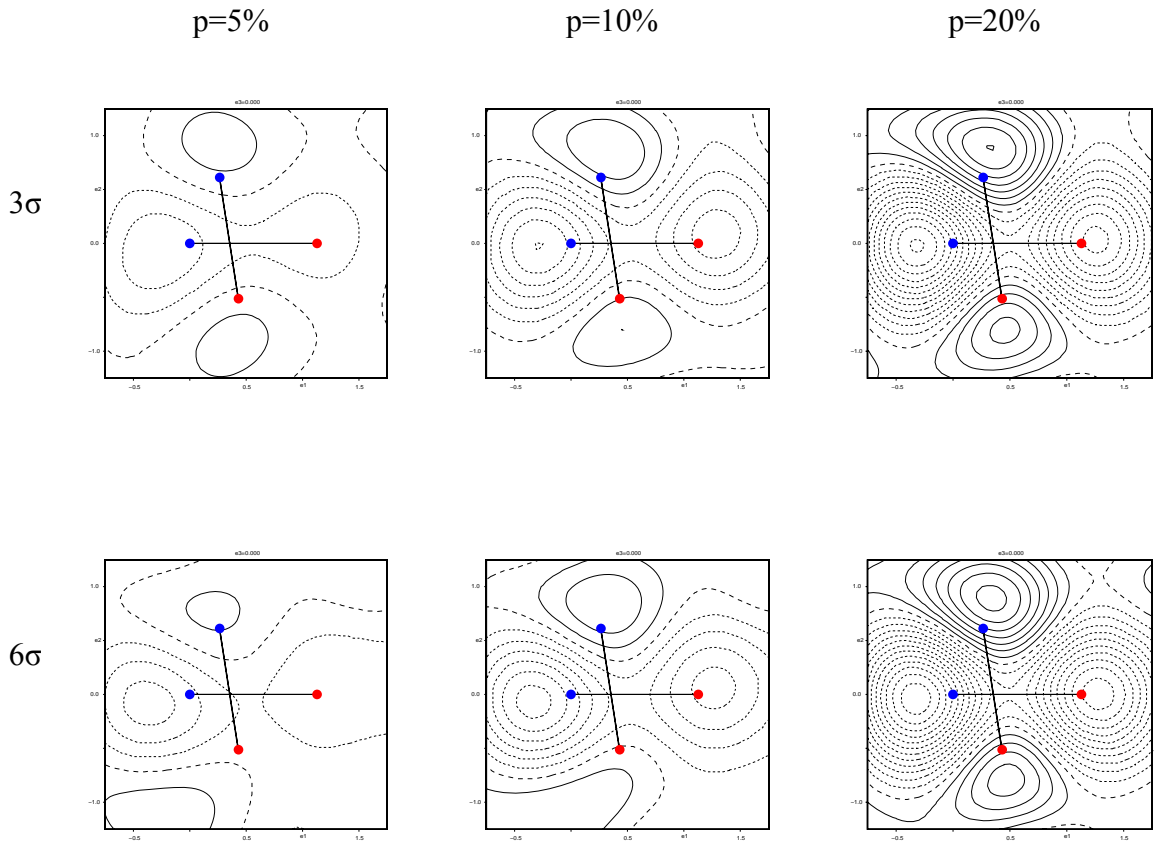

* Correspondence author (Jurg.Schefer@psi.ch)    

**Fig 6b**

b) sin(Θ)/λ up to 0.7 Å$^{-1}$

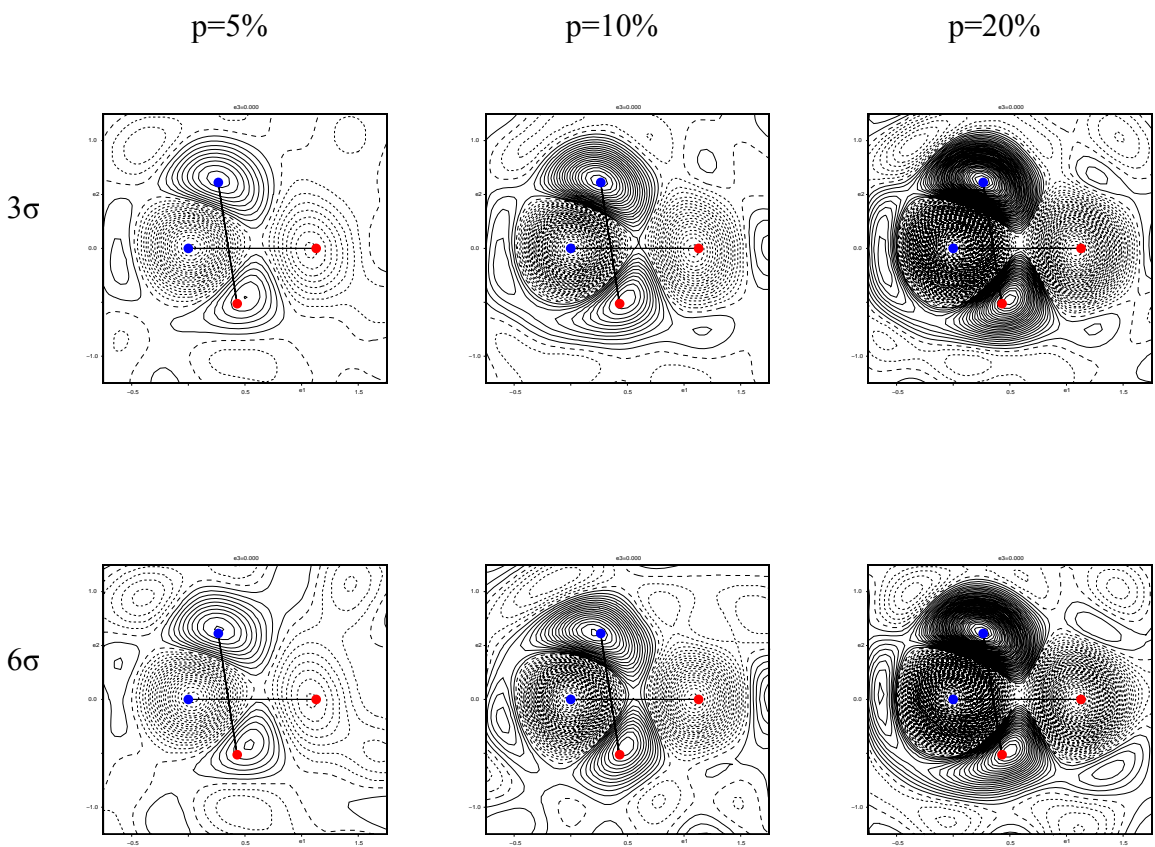

**Figure 6:** Difference Photo Fourier maps in the NO-planes of the ground state GS and the side-on configuration of the metastable state SII of Sodiumnitroprusside (SNP). The maps correspond to expected Fourier maps for different completeness of the data sets (sin(Θ)/λ < 0.4 Å$^{-1}$(a) and 0.7 Å$^{-1}$ (b), different populations, p, and different counting statistics, σ. The vertical NO bond corresponds to the ground state, the horizontal one to the side-on configuration of SII. The lower and the right atom in each figure is oxygen (online version: red), the top at left atom nitrogen (online version: blue). All figures display the same section defined by the atoms N4(GS), O1(GS) and the side on configuration of N4' (SII) and represent an area of 2.5 Å by 2.5 Å. The crystallographic c axis of SNP is therefore practically vertical to the plane displayed.

# Acknowledgement

Neutron beam time on the instruments TriCS and Morpheus of the Laboratory for Neutron Scattering (LNS, ETH Zurich and Paul Scherrer Institute) at the Swiss Spallation Neutron Source SINQ, Switzerland is gratefully acknowledged. The development of the JANA2000 and JANA2006 program packages was supported by the Grant Agency of the Czech Republic, grant 202/06/0757. Furthermore, financial support by the DFG (WO618/5-3) is gratefully acknowledged.

# References

[1] Woike, Th.; Krasser, W.; Bechthold, P.S.;
Extremely long living metastable state of $Na_2[Fe(CN)_5NO] \cdot 2H_2O$ single crystals: Optical properties.
*Phys. Rev. Lett*. **53** (1984) 1767-1770

[2] Decurtins, S.; Gütlich, P.; Köhler, C.P.; Spiering, H.; Hauser, A.:
Light-induced excited spin state trapping in a transition-metal complex: the hexa-1-propyltetrazole-iron (II) tetrafluoroborate spin-crossover system.
*Chem. Phys. Lett.* **105** (1984) 1-4

[3] Gütlich, P.; Garcia, Y.; Woike, Th.:
Photoswitchable coordination compounds.
*Coord. Chem. Rev*. **219-221** (2001) 839-879

[4] Bitterwolf, Th.E.:
Photochemical nitrosyl linkage isormerism/metastable states.
*Coordination Chemistry Reviews* **250** (2006) 1196-1207

[5] Coppens, P.; Novozhilova, I.; Kovalevsky, A.:
Photoinduced linkage isomers of transition-metal nitrosyl compounds and related complexes.
*Chem. Rev.* **102** (2002) 861-883

[6] Rüdlinger, M.; Schefer, J.; Vogt, Th.; Woike, Th.; Haussühl, S.; Zöllner, H.:
Ground- and light induced metastable states of sodiumnitroprusside.
*Physica B* **180&181** (1992) 293-298

[7] Schaniel, D.; Imlau, M.; Weisemoeller, Th.; Woike, Th., Krämer, K. W., Güdel, H.-U.:
Photoinduced nitrosyl linkage isomers uncover a variety of unconventional photorefractive media.
*Advanced Materials* **19**, (2007) 723–726

[8] Krasser, W.; Woike Th.; Haussühl S.; Kuhl, J.; Breitenschwerth, A.:
Resonance Raman Spectroscopy from the electronic metastable state of an $Na_2[Fe(CN)_5NO] \cdot 2H_2O$ single crystal.
*J. Raman Spectr*. **17** (1986) 83-87

[9] Pressprich, M.R.; White, M.A.; Vekhter, Y.; Coppens, P.:
Analysis of a metastable electronic excited state of sodium nitroprusside by X-ray crystallography.
*J. Am. Chem. Soc.* **116** (1994) 5233-5238

[10] H.U. Güdel:
Comment on the nature of light-induced metastable states in nitroprussides.
*Chem. Phys. Lett*. **175** (1990) 262-266

[11] Carducci, M.D.; Pressprich, M.R.; Coppens, P.:
Diffraction studies of photoexcited crystals: Metastable nitrosyl-linkage isomers of sodium nitroprusside.
*J. Am. Chem. Soc*. **119** (1997) 2669-2678

[12] Delley, B.; Schefer, J.; Woike, Th.:
Giant lifetimes of optically excited states and the elusive structure of sodiumnitroprusside.
*J. Chem. Phys.* **107** (1997) 10067-10074.

[13] Schaniel, D.; Schefer, J.; Imlau, M.; Woike, Th.:
Light-induced structural changes by excitation of metastable states in $Na_2[Fe(CN)_5NO]\cdot 2H_2O$ single crystals.
*Phys. Rev. B* **68** (2003) 104108, 1-11

[14] Schaniel, D.; Schefer, J.; Imlau, M.; Woike, Th.:
Structure of the light-induced metastable state SII in $Na_2[Fe(CN)_5NO]\cdot 2H_2O$.
*Phys. Rev. B* **71** (2005) 174112,1-7

[15] Schaniel, D.; Woike, Th.; Schefer, J.; Petříček, V.; Krämer, K. W.; Güdel, H. U.:
Neutron diffraction shows a photoinduced isonitrosyl linkage isomer in the metastable state *S*I of $Na_2[Fe(CN)_5NO]\cdot 2D_2O$.
*Physical Review B* **73**, (2006) 174108, 1-5

[16] Petříček, V.; Dusek M.; Palatinus, L.:
The crystallographic computing system JANA2000.
Institute of Physics, Praha (2000)

[17] Schefer, J.; Könnecke, M.; Murasik, A.; Czopnik, A.; Strässle, Th.; Keller, P.; Schlumpf, N.:
Single-crystal diffraction instrument TriCS at SINQ.
*Physica B* **276-278** (2000) 168-169

[18] Fischer, W.E.:
SINQ - The spallation neutron source, a new research facility at PSI.
*Physica B* **234-236** (1997) 1202-1208

[19] Zanini, L.:
Summary Report for MEGAPIE R&D Task Group X9: Neutronic and Nuclear Assessment.
*PSI Bericht* Nr. **5-12**, ISSN 1019-0643 (2005)

[20] Schefer, J.; Woike, Th.; Haussühl, S.; Fernandez Díaz, M. T.:
Population and structural changes of the metastable state II in sodiumnitroprusside $Na_2[Fe(CN)_5 NO]\cdot 2H_2O$ at 60K.
*Zeitschrift für Kristallographie* **212** (1997) 29-33

[21]    Schaniel, D.;
        Structural investigations of high-knowledge-content materials.
        *Thesis*, No. **14902**, ETH Zurich (2002)
[22]    Woike, Th.; Krasser, W.; Zöllner, H.; Kirchner, W.; Haussühl, S.:
        Population dynamics of the two light induced metastable states in
        $Na_2[Fe(CN)_5NO] \cdot 2H_2O$ single crystals.
        *Z. Phys. D* **25** (1993) 351-356
[23]    Krasser, W.; Woike Th.; Bechthold , P.S; Haussuehl, S.:
        Raman Spectra of $Na_2[Fe(CN)_5 NO] \cdot 2H_2O$ in the long living metastable
        electronic state.
        *J. Molecular Structure,* **114** (1984) 57-60
[24]    Woike ,Th.; Zöllner, H.; Krasser, W..; Haussühl, S.;
        Raman-spectroscopy and differential scanning calorimetric studies of the light
        induced metastable states in $K_2(RuCl_5)NO$.
        *Solid State Commun.* **73** (1990) 149-152
[25]    Schaniel, D.; Schefer, J.; Delley, D.; Imlau, M.; Woike, Th.:
        Light-induced absorption changes by excitation of metastable states in
        $Na_2[Fe(CN)_5NO] \cdot 2H_2O$ single crystals.
        *Phys. Rev. B* **66**, (2002) 085103,1-10
[26]    Woike, Th.; Imlau, M.; Angelov, V.; Schefer, J.; Delley, B.:
        Angle dependent Mössbauerspectroscopy in the ground and metastable
        electronic state SI in $Na_2 [Fe(CN)_5NO] \cdot 2H_2O$ single crystals.
        *Phys. Rev. B*, **61** (2000) 12249-12260
[27]    Woike, Th.; Kirchner, W.; Kim, Hyung-sang; Haussühl, S; Rusanov, V.;
        Angelov, V.; Ormandjiev, S.; Bonchev, Ts.:
        Mössbauer parameters of the two long-lived metastable states in
        $Na_2[Fe(CN)_5NO]2H_2O$ single crystals.
        *Hyperfine Interactions*, **77** (1993) 265-275